\documentclass[aip,reprint]{revtex4-1}
\usepackage{graphicx}
\usepackage{dcolumn}
\usepackage{amsmath}
\begin{document}

\title{Liquid crystal hyperbolic metamaterial for wide-angle negative-positive refraction and reflection}

\author{G. Pawlik}
\email{grzegorz.pawlik@pwr.wroc.pl}
\affiliation{Institute of
Physics, Wroclaw University of Technology, Poland}

\author{K. Tarnowski}
\affiliation{Institute of Physics, Wroclaw University of
Technology, Poland}

\author{W. Walasik}
\affiliation{Aix Marseille Universite, Institut Fresnel, France}

\author{A.C. Mitus}
\affiliation{Institute of Physics, Wroclaw University of
Technology, Poland}

\author{I.C. Khoo}
\affiliation{Department of Electrical Engineering, Pennsylvania
State University,  USA}

\begin{abstract}
We show  that  nanosphere dispersed liquid crystal (NDLC)
metamaterial can be characterized in near IR spectral region as an
indefinite medium  whose real parts of effective ordinary and
extraordinary permittivities  are opposite in signs. Based on this
fact we design a novel electrooptic effect: external electric
field driven switch between normal refraction, negative refraction
and reflection of TM incident electromagnetic wave from the
boundary vacuum/NDLC. A detailed analysis of its functionality is
given based on effective medium theory combined with a study of
negative refraction in anisotropic metamaterials, and Finite
Elements simulations.
\end{abstract}

\maketitle

\section{Introduction}

Novel metamaterials \cite{1,15}, in particular those for which the
spatial distribution of optical parameters can be specifically
tailor--made, have proven to be a viable route to realize various
linear, nonlinear or tunable optical properties and processes.
Amongst them, the hyperbolic metamaterials (HM)  - indefinite
metamaterials, for which the permittivity and permeability tensors
are negative along only certain of the principal axes of the
metamaterial \cite{SmithS1} - attract interest because of
possibilities of optical steering and manipulation. Closely
related are optical effects like negative refraction
\cite{SmithJosa2004,FangKS1}, hyperlensing
\cite{JacobAN1,LeeLXZS1,YaoTWLBWZ1,CasseLHGMS1}, cancellation of
reflection and transmission \cite{YangHLLJZ1}. Optical devices
based on indefinite media are designed, like, e.g., polarization
beam splitters \cite{ZhaoCF1}, angular filters
\cite{AlekseyevNTLBN1}, optical transmission modulator
\cite{switch}. Theoretical approaches were worked out to handle
the negative refraction in anisotropic indefinite media have also
been presented \cite{SmithJosa2004,Grzegorczyk,Liu}.

Tunability and simple production become important requirements.
One of the few metamaterials which offer tunability \cite{JOSAB},
is nematic liquid crystal (NLC) doped with coated core-shell
spheres (NDLC) \cite{16}. Recently we have demonstrated a negative
refraction in near infra-red (IR) in this system with a planar NLC
configuration for a wide interval of incident angles of
electromagnetic (EM) wave, using finite elements (FE) calculations
(Comsol) \cite{SPIE12}. Similar results in visible light range
were presented in Ref. \cite{Jia2} where it was shown that a HM
consisting of NLC doped with a sufficiently high concentration of
silver nanoparticles can become a HM; all-angle negative
refraction for homeotropic NLC orientation was demonstrated based
on geometric arguments.

The point we would like to emphasize here is that a medium is
indefinite is neither necessary \cite{negrefrLC2,negrefrLC1} nor a
sufficient condition for negative refraction; we shall show below
that for a HM a positive refraction as well as reflection can take
place. Thus, additional quantitative calculations are necessary to
trace the path of EM in an indefinite medium.

The aim of this paper is to design, using effective medium
approximation and FE calculations, an electric field driven device
for switching in  NDLC HM between three scenarios:  negative
refraction, positive refraction and reflection for wide intervals
of  angles of incidence of EM and of NLC director orientations. To
this end, the approach based on negative refraction in anisotropic
metamaterials  \cite{Grzegorczyk,Liu} is used.

\section{NDLC as a tunable indefinite medium}

Consider a rectangular NLC cell with thickness $L$ along $x$
direction, filled with NDLC metamaterial described below, with
planar alignment of NLC molecules in $x-y$ plane (Fig.
\ref{model}(a)). The incident light with transverse magnetic (TM)
polarization ($\vec{{E}}$ in $x-y$ plane and $\vec{{B}}$ along $z$
direction) propagates in $x-y$ plane and impinges  as an
extraordinary wave onto the NLC host. The permittivity tensor for
NLC reads \cite{Werner1}:

\begin{equation}
\varepsilon_{LC} = \left( \begin{array}{ccc}
\varepsilon_\bot + \Delta \varepsilon \cos^2 \gamma    & \Delta \varepsilon \sin \gamma \cos \gamma  & 0 \\
\Delta \varepsilon \sin \gamma \cos \gamma             & \varepsilon_\bot + \Delta \varepsilon \sin^2 \gamma & 0 \\
0                                                      & 0
&\varepsilon_\bot
\end{array} \right),
\end{equation}

\noindent where $\gamma$ denotes the angle between the $+x$ axis
and the  director $\vec n$, $\Delta \varepsilon=
\varepsilon_{\parallel}-\varepsilon_{\bot }$,
$\varepsilon_{\parallel}=n_e^2$, $\varepsilon_\bot=n_o^2$.

The NDLC metamaterial \cite{16} comprises  the host nematic liquid
crystal containing uniformly distributed  non--magnetic spheres
with core radius $r_{1}$, made of polaritonic material and
semiconductor shell (Drude material) with thickness $d$. The
effective permittivity and permeability of NDLC were calculated in
Ref. \cite{16} using Mie theory  and Maxwell Garnet mixing rule:
\begin{equation}
\label{eq} \varepsilon_{^{eff}} =\varepsilon_3 \Bigg(\frac{k_3^3
+4\pi iNa_1 }{k_3^3 -2\pi iNa_1 }\Bigg),\,\,\mu_{^{eff}}
=\frac{k_3^3 +4\pi iNb_1 }{k_3^3 -2\pi iNb_1 },
\end{equation}
where $k_{3} =\sqrt {\varepsilon_{3} } \,2\pi /\lambda $, $\lambda
$ denotes the free--space wavelength, $\varepsilon_{3} \quad -$
permittivity of the NLC host along main axes, $N$ -- number
density of the spheres, and $a_{1}, b_{1} $ -- scattering
coefficients. Formula (\ref{eq}) is valid in long-wavelength limit
for low values of filling factor $f=4\pi (r_{1} +d)^{3}N/3$.
Parameters used in this paper are: $r_1=0.05\,\rm{\mu m}$,
$d=0.01\,\rm{\mu m}$, $f=0.06$, $\varepsilon(\infty)=17$,
$\omega_T/2 \pi= 360\, \rm{THz}$, $\omega_L/2 \pi= 855\,
\rm{THz}$, $\gamma_1/2 \pi= 3.75\, \rm{THz}$, $\omega_p \pi= 230\,
\rm{THz}$, $\gamma_2=\omega_p/60$, $\lambda=1.75\, \rm{\mu m}$,
see Ref. \cite{16} for the definitions. Plots of the  real parts
$\mathop {{\varepsilon }'}\nolimits_{eff} = \Re\,\{{\mathop
{{\varepsilon }}\nolimits_{eff}}\}$, $\mathop {{\mu
}'}\nolimits_{eff} = \Re\,\{{\mathop {{\mu }}\nolimits_{eff}}\}$
in function of $\varepsilon_3$ are shown in Fig. \ref{model}(b).


\begin{figure}[htb]
\includegraphics[scale=0.8]{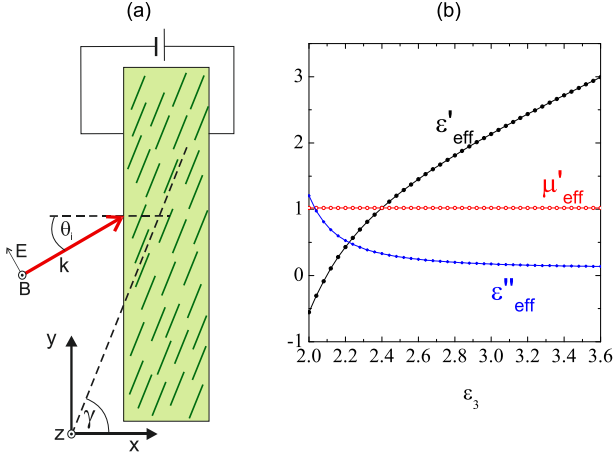}
\caption{\label{model}(Color online) (a) Geometry of the system.
(b) Dependence of effective permittivity and permeability on
permittivity of the LC host $\varepsilon_{3}$, Eq. (\ref{eq}).}
\label{model}
\end{figure}

The propagation of a TM incident wave in this medium  can be
described by specifying $\mathop {{\varepsilon }'}\nolimits_{eff}$
in the propagation plane along principal axes and $(\mathop {{\mu
}'}\nolimits_{eff} )_{zz}$ \cite{SmithJosa2004}. Those components
were calculated from Eq. (\ref{eq})  using the tensor components
of $\varepsilon_{3}$ along main axes: $\varepsilon_{\parallel}$
and $\varepsilon_{\bot }$ \cite{OptLett}. In this paper we put
$\varepsilon_{\bot }=2, \varepsilon_{\parallel}=2.91$. Then,
$(\mathop {{\varepsilon }'}\nolimits_{eff} )_{\bot }\simeq -0.52$
for $\varepsilon_{3}=\varepsilon_{\bot }$ and $(\mathop
{{\varepsilon }'}\nolimits_{eff} )_{\parallel }\simeq 2$ for
$\varepsilon_{3}=\varepsilon_{\parallel}$. As the real parts of
effective permittivities along principal axes are opposite in
signs, NDLC is an indefinite medium.  The effective permeability
is independent on $\varepsilon_{3}$ in the interval of values
relevant for this design (Fig. \ref{model}(b)) and we put
$(\mathop {{\mu }'}\nolimits_{eff} )_{zz} \equiv \mathop {{\mu
}'}\nolimits_{eff} \simeq 1$.

\begin{figure}[htb]
\includegraphics[scale=0.65]{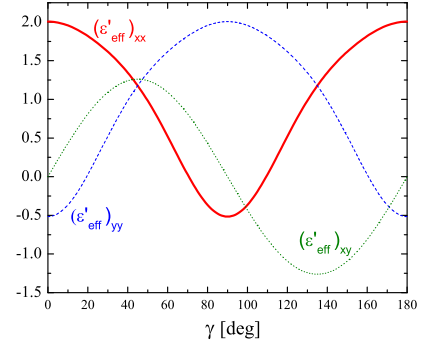}
\caption{The dependence of tensor components $(\mathop
{{\varepsilon }'}\nolimits_{eff} )_{xx}$, $(\mathop {{\varepsilon
}'}\nolimits_{eff} )_{yy}$ and  $(\mathop {{\varepsilon
}'}\nolimits_{eff} )_{xy}$  on rotation angle $\gamma$.}
\label{rotation}
\end{figure}

The tensor components $(\mathop {{\varepsilon }'}\nolimits_{eff}
)_{xx}, (\mathop {{\varepsilon }'}\nolimits_{eff} )_{yy}, (\mathop
{{\varepsilon }'}\nolimits_{eff} )_{xy}$  in coordinates from Fig.
\ref{model}(a), calculated by rotating  the diagonal tensor in
principal axes around $z$ axis by angle $\gamma$, are shown in
Fig. \ref{rotation}. As angle $\gamma$ characterizes  the
direction of the optic axis which can be varied through the
external electric field, we conclude that the NDLC becomes (within
the limits of validity of the effective medium theory) a tunable
hyperbolic medium.

\section{Refraction on the boundary isotropic/indefinite material.
All-angle negative refraction}

\begin{figure}[htb]
\includegraphics[scale=0.77]{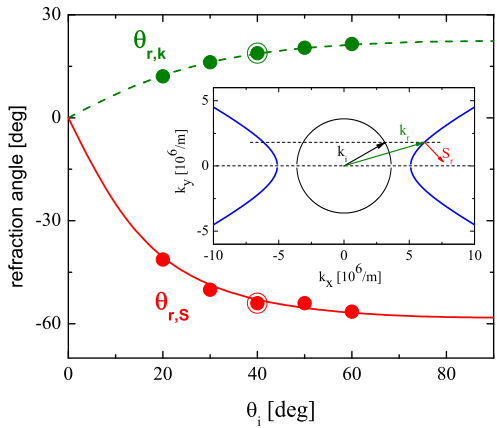}
\caption{\label{all} (Color online) Plots of refraction angles
$\theta_{r,S}(\theta_i)$ (solid line) and $\theta_{r,k}(\theta_i)$
(dashed line). Symbols represent FE results \cite{SPIE12}. Large
symbol - see Fig. \ref{gamm40}. Inset: isofrequency contours for
the TM wave incident from an isotropic material to an indefinite
one with $(\mathop {{\varepsilon }'}\nolimits_{eff} )_{xx}<0$ and
$(\mathop {{\varepsilon }'}\nolimits_{eff} )_{yy}>0$ for the angle
of incidence  $\theta_i=30^\circ$.}
\end{figure}

Rotation of the director of NLC accompanied by the variation  of
the effective permittivity tensor  changes  the refraction at the
boundary isotropic/indefinite medium. In geometric approach this
rotation promotes the rotation of the hyperbola of equifrequency
contour in NDLC material \cite{Grzegorczyk}. In the principal axes
of effective permittivity tensor this hyperbola determines the
dispersion relation for TM waves in NDLC \cite{15}:

\begin{equation}\label{disp}
    \frac{k_x^2}{(\mathop
{{\varepsilon }'}\nolimits_{eff} )_{yy}}+\frac{k_y^2}{(\mathop
{{\varepsilon }'}\nolimits_{eff} )_{xx}}= \frac{\omega^2}{c^2},
\end{equation}

\noindent where $c$ denotes the velocity of light  in vacuum. The
case of $\gamma=90^\circ$ is shown as inset in Fig. \ref{all},
where the equifrequency contour of an isotropic material is
represented by a circle. Based on this diagram, the geometrical
analysis of refraction can be done in a standard way
\cite{Grzegorczyk} by observing the continuity of the tangential
components of incident ($\vec k_i$) and refracted ($\vec k_r$)
wave vectors. The direction of energy flow described by the
Poynting vector $\vec{S}$ within the indefinite medium is
indicated by the arrow drawn normal to the hyperbolic isofrequency
surface. The refraction angles for the wave beam $\theta_{r,S}$
and for the wave vector $\theta_{r,k}$ can be calculated
analytically. The same geometric approach is used in the case of
rotated nematic/equifrequency contour (see, e.g., appendix in Ref.
\cite{Grzegorczyk}).

\begin{figure}[htb]
\includegraphics[scale=0.32]{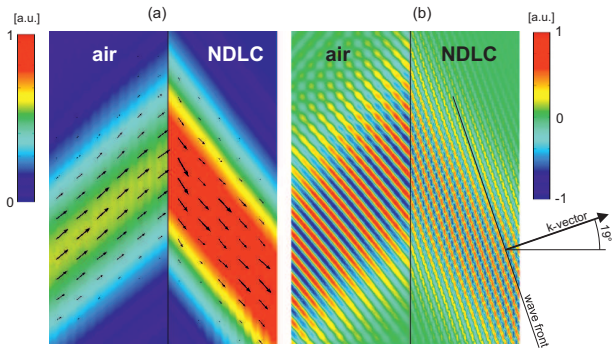}
\caption{\label{gamm40} (Color online) Finite Element simulations
of the electric and magnetic field mapping for gaussian beam with
TM polarization, $\theta_i=40^\circ$. Magnitude of the electric
field and Poynting vectors (a), $z$-component of magnetic field
(b).}
\end{figure}

In the case of  uniform configuration of NLC ($\gamma=90^\circ$)
the geometric approach predicts that negative refraction of the TM
wave occurs for all angles of incidence $\theta_i$ (solid line in
Fig. \ref{all}). This prediction is confirmed by FE calculations
(symbols in Fig. \ref{all}). Fig. \ref{gamm40} characterizes the
energy flow and electromagnetic wave in the system for angle of
incidence $\theta_i=40^\circ$. Part (a) shows the magnitude of the
electric field  and Poynting vectors, part (b) -- the
$z$-component of magnetic field. Direction of vector $\vec k_r$
for refracted wave and angle of negative refraction for Poynting
vector are in agreement with analytical results from Fig.
\ref{all}.

\section{Electric switch between negative/positive refraction and
reflection}

\begin{figure}[htbp]
\includegraphics[scale=0.76]{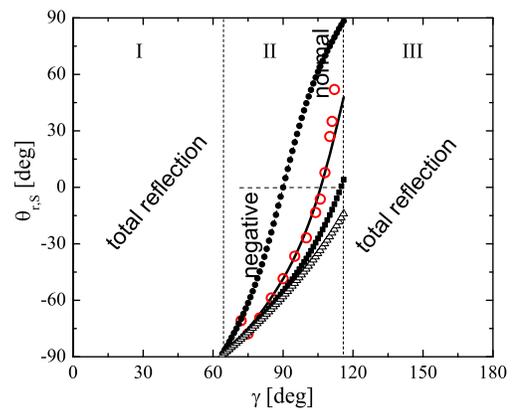}
\caption{\label{diagram} Refraction angle $\theta_{r,S}$ in
function of the orientation $\gamma$ of NLC director  for
$\theta_i=0^\circ$ (full circles), $30^\circ$ (solid line),
$50^\circ$ (squares) and $80^\circ$ (triangles). Large circles
represent FE calculations for $\theta_i=30^\circ$.}
\end{figure}

Using the formalism presented in the  last Section we have studied
the behaviour of a plane TM wave incident from vacuum on the
indefinite NDLC for a few chosen angles of incidence $\theta_i$
and for an arbitrary orientation of NLC director ($0 \leq \gamma
\leq 180^\circ$). The dependence of refraction angle
$\theta_{r,S}$ on $\gamma$ is shown in Fig. \ref{diagram}. We have
singled out three distinct regions. We first discuss Region II
that was calculated using analytical geometric approach and
extends from $\gamma =\gamma_N\simeq 64^\circ$ to $\gamma =
\gamma_P \approx 116^\circ$. The location of its boundaries is
independent  on $\theta_i$. The negative refraction for the
Poynting vector is present for all angles of incidence $\theta_i$,
while positive refraction requires that $\theta_i < \theta_{i, 0}
\approx 50^\circ$. For example, for $\theta_i=30^\circ$ negative
refraction is present for $\gamma_N < \gamma_0 \approx 106^\circ$,
positive refraction -- for $\gamma > \gamma_0$. We have used  the
approach from Ref. \cite{switch} to show that in this region the
wave vector of the wave propagating in NDLC HM metamaterial is
purely real which implies that the plane wave impinging from
vacuum excites an homogeneous wave. In regions I ($\gamma <
\gamma_N$) and III ($\gamma > \gamma_P$) the geometric
construction breaks down, the wave vector becomes imaginary, which
implies that the impinging plane wave excites an evanescent wave.

To verify analytical geometric predictions we have used FE
calculations. In Regions I and III total reflection occurs (there
was no transmitted wave) while in Region  II the simulations for
$\theta_i=30^\circ$ (symbols) confirm theoretical analysis. Small
discrepancies close to boundaries of Region II have numerical
origin.

\begin{figure}[htbp]
\includegraphics[scale=0.5]{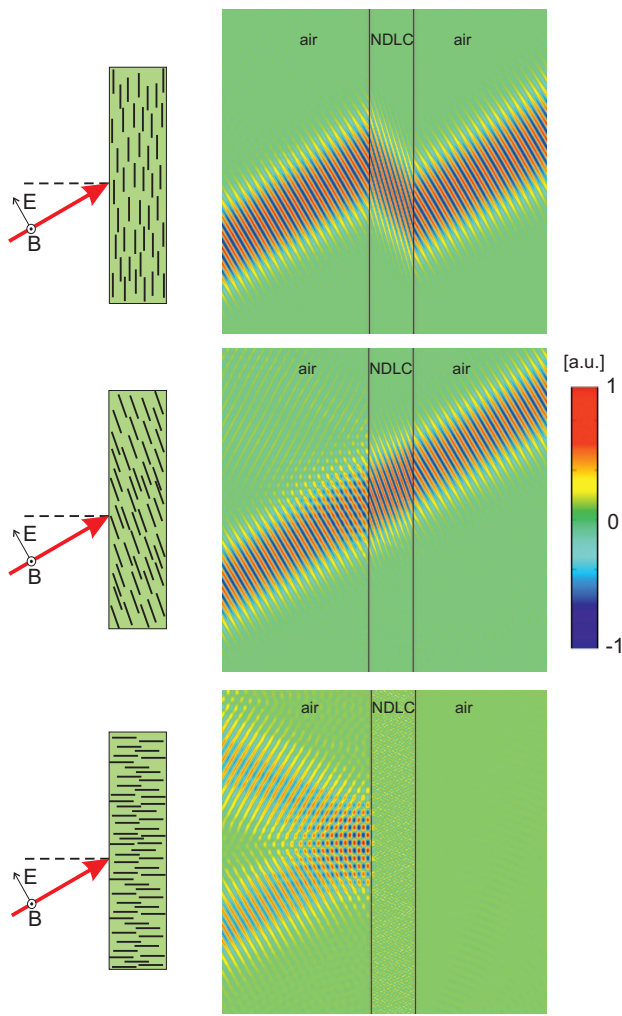}
\caption{\label{FE} (Color online) Operating principle of the
switch. Left: orientation of NLC host, right: magnetic field
mapping for a gaussian TM beam incident at $\theta_i=30^\circ$.
Negative refraction ($\gamma=90^\circ$, top); positive refraction
($\gamma=110^\circ$, middle); total reflection ($\gamma=0^\circ$,
bottom).}
\end{figure}

These results make possible a design of a novel electrooptic
effect: external electric field driven real time switch between
negative refraction, positive refraction and total reflection. To
illustrate its functionality we have performed FE simulations of a
TM gaussian beam with $\lambda_0=1.74\,\mu$m impinging a NDLC HM
cell with thickness $10\,\mu$m at angle of incidence
$\theta_i=30^\circ$. In Fig. \ref{FE} the magnetic field mapping
is shown for three operating states of the switch, which depend on
the orientation of the NLC host controlled by external electric
field: negative refraction for $\gamma=90^\circ$ (top); positive
refraction for $\gamma=110^\circ$ (middle) and total reflection
for $\gamma=0^\circ$ (bottom).

\section{Conclusions}

We have shown using effective medium approach and FE calculations
(Comsol) that nanosphere dispersed liquid crystal metamaterial
becomes a tunable indefinite medium in near IR. The tunability is
provided by varying the external electric field which changes the
orientation of the NLC director, the overall effective
permittivity tensor and  the orientation of the hyperbola of the
dispersion relation of extraordinary waves through the medium. The
latter allows to control the character of electromagnetic field in
the medium: propagating or evanescent. We have exploited this
tunability to design a new electrooptic effect for the
manipulation of light: electric field driven switch between
negative and positive refraction and reflection, and have
illustrated its functionality. Compared to negative refraction in
pure NLC medium \cite{negrefrLC2,negrefrLC1} the NDLC device
operates in  much wider intervals of incidence/refraction angles.

\textbf{Acknowledgements}

This work is supported by the Air Force Office of Scientific
Research (AFOSR). G.P. thanks Polish National Science Centre for
financial support under Grant NN507 322440. K.T. acknowledges
support of the Foundation for Polish Science START Program.

\end{document}